# A GPP ALGORITHM FOR HIPPOCAMPAL INTERNEURON CHARACTERIZATION

KATHERINE MEDINA

*SCHOOL OF SCIENCE AND TECHNOLOGY, UNIVERSITY OF VALPARAÍSO, VALPARAÍSO, REGIÓN DE VALPARAÍSO, CHILE*

*CORRESPONDING AUTHOR: KATHERINE MEDINA (KM2594@EMAIL.VCCS.EDU)*

## Abstract

Correctly identifying neuronal subsets is critical to multiple downstream methods in several areas of neuroscience research. The hippocampal interneuron characterization technology has achieved rapid development in recent years. However, capturing true neuronal features for accurate interneuron characterization and segmentation has remained elusive. In the current study, a novel global preserving estimate algorithm is used to capture the nonlinearity in the features of hippocampal interneurons after factor Algorithm. Our results provide evidence for the effective integration of the original linear and nonlinear neuronal features and achieves better characterization performance on multiple hippocampal interneuron databases through Array matching.

## Key words

Hippocampal Interneuron Characterization, Globally Preserving Estimate, Factor Algorithm

## 1. INTRODUCTION

As a key area of biologic characterization technology, hippocampal interneuron characterization technology has achieved rapid development in recent years. A large number of Array characterization technologies are applied in this field. The main application areas include: FACS based characterization, immunological characterization, man-machine interaction and so on.

From the main process of hippocampal interneuron characterization we can see that the feature mining and processing as well as classifier selection are the main factors that affect the characterization performance. The Factor Algorithm (FA) [1], Linear Differential Algorithm (LDA) [2] and Globally Preserving Estimates (GPP) [3] algorithm have been proved to be able to greatly improve the characterization rate. It is found that the hippocampal interneuron global feature mining can represent the hippocampal interneuron overall contour and any other global data, but it is sensitive to changes of illumination and posture. The native feature mining can represent the native detail of hippocampal interneuron and has certain robustness to light, but representational capacity for overall data is not strong [4]. The union of global eigen hippocampal interneurons and native Gabor features improves the characterization rate, but due to the complexity of computation and the long computation time, its application scope is limited [5].By using the TAT feature to describe the hippocampal interneuron features, and extracts the global data and native data. Characterization ability is improved to some extent. However, the union process of global features and the native features is just amalgamating, lacking effective integration [6].

In this paper, aiming at the shortcomings of the algorithm, the GPP is introduced to extract the features. At the same time, Euclidean correlation Algorithm [7] is used to integrate the global features and the native features effectively, and constructs classifier by the Array Matching algorithm [8]. After comparing and aggregating multiple hippocampal interneuron databases, the algorithm is better than contrast algorithm, and has high characterization performance.

## 2. FEATURE MINING AND DIMENSION DECLINE ALGORITHM

### 2.1 TAT FEATURE MINING

TAT feature descriptor is evolved from the HULT operator, which can describe the native difference data of the imprint well and is not easily disturbed by the noise. The TAT feature mining process is as follows:

Step1. Imprint cropping, in this paper, the imprint length and height can be divisible by 16;

Step2. The divided sample imprint is a number of small blocks, each consists of 4 adjacent cells. The block slides in the form of overlapping two cells, selects unsigned gradient direction, the number of the gradient direction angles is 9, the parallel and vertical gradients of the pixel point $(x, y)$ in the imprint are computed as follows:

$$I_x(x, y) = I(x+1, y) - I(x-1, y) \qquad (1)$$

$$I_y(x, y) = I(x, y+1) - I(x, y-1) \qquad (2)$$

The gradient amplitude and the gradient direction of the pixel point $(x, y)$ are derived respectively

$$m(x, y) = \sqrt{(I_x(x, y))^2 + (I_y(x, y))^2} \qquad (3)$$

$$\theta(x, y) = \arctan\left(\frac{I_x(x, y)}{I_y(x, y)}\right) \qquad (4)$$

The histogram features of all the pixels in all directions are counted so as to obtain the histogram features of each cell, and further the histogram of each block is derived.

Step3. Cascading the histogram features of each block sequentially to obtain the overall TAT feature $S_A$ of the overall imprint.

For instance: an imprint with a width of 92 * height of 112, each line has (112-16) / 8 + 1 = 13 blocks, since each column cannot be divisible by 16, first incise it to make it into 80, and then compute the total (80-16) / 8 +1 = 9 blocks, then the whole imprint has 13 * 9 = 117 blocks, each of which has 2 * 2 cells and each cell has 9 bins, and all the feature dimensions of each block are 2 * 2 * 9 = 36 dimensions, thus the TAT feature dimensions of the entire imprint is 117 * 36 = 4212.

In the selection of native features, the difference of the coordinates of the key parts of hippocampal interneuron is relatively small. After comparison, the selected $L$ pcs sub-blocks can contain most of the classified data easily, which can not only reduce the feature dimension and computational complexity, but also the characterization rate will not change too much. In this paper, it adopts the same steps as the global feature mining method to extract the above $L$ pcs sub-block features. The original native TAT features are $\{B_i | i = 1, 2, \text{L}, L\}$.

## 2.2 GPP ALGORITHM

Global features and native features derived through TAT feature mining still have certain complexity in describing data. In order to improve characterization performance, some transpositions are needed to extract features. For global features, 2DFA algorithm is used to reduce dimensionality, compared with the traditional FA, 2DFA algorithm is easier and more accurate to compute distribution Array, and its corresponding eigenvector is faster; for native features, this paper uses the traditional FA dimensionality decline method. Because the essence of FA and 2DFA algorithms are still linear, it cannot effectively describe the non-linear change of hippocampal interneuron imprint data. In this paper, Globally Preserving Estimates (GPP) algorithm is applied to non-linear data in hippocampal interneuron imprints capture, GPP algorithm is a kind of new subspace Algorithm method, a linear approximation of the Laplacian Eigen map nonlinear method, which not only solves the shortcomings of the traditional linear methods such as FA that is difficult to maintain the original data nonlinear manifold, but also solves the shortcomings of nonlinear method that is difficult to obtain the new sample point's low-dimensional estimate.

Globally preserving estimates look for linear transpositions $W$ through certain performance goals in order to achieve dimensionality decline for high-dimensional data:

$$y_i = W^T x_j, j = 1, 2, \text{L } l \quad (5)$$

Given the existence of $l$ training segments $X = \{x_i\}_{i=1}^{l} \in R^m$, the transposition Array can be derived by minimizing the following objective function:

$$\min(\sum_{i,j} W^T x_i - W^T x_j)^2 S_{ij}) \quad (6)$$

Among them, $S$ is the weight Array, which can be defined by $k$ nearest neighbor method:

$$S_{ij} = \frac{\exp(\|x_i - x_j\|^2)}{t} \quad (7)$$

Among them, $x_j$ is the $k$ nearest neighbor of $x_i$

Conduct formula (6) algebraic changes

$$\begin{aligned} &\tfrac{1}{2} \sum_{i,j} W^T x_i - W^T x_j)^2 S_{ij} \\ &= \sum_{i,j} W^T x_i D_{ii} x_i^T W - \sum_{i,j} W^T x_i S_{ij} x_i^T W \\ &= W^T X (D-S) X^T W \\ &= W^T X L X^T W \end{aligned} \quad (8)$$

Among them, $X = [x_1, x_2, \text{L } x_l]$, $D$ is a $l \times l$ diagonal Array, diagonal element is $D_{ii} = \sum_j S_{ij}$, $L = D - S$.

The transposition Array $W$ deduced minimum value can be derived by solving the following generalized eigenvalue problem.

$$XLX^T W = \lambda XDX^T W \quad (9)$$

The eigenvectors corresponding to $d$ pcs smallest non-zero eigenvalues of the above formula form a estimate Array $W = [w_1, w_2, \text{L } w_d]$. Due to the small sample problem existing in the GPP algorithm, 2DFA and FA are used to reduce the dimension of the global feature and the native feature respectively, then the GPP algorithm is applied to construct the facial features in the reduced dimension subspace. In order to be able to perform 2DFA processing on the global TAT feature, the global TAT feature vector is considered as an array, and the final global TAT feature $F_A$ is derived finally through dimensionality decline by the above algorithm.

For native features $B_i$, the FA algorithm is firstly used to reduce the dimension and then GPP algorithm is used to obtain the final native TAT feature $\{F_i | i = 1, 2, \text{L}, L\}$ in the FA subspace.

## 3. FEATURE UNION AND CLASSIFIER LEARNING ALGORITHM

### 3.1 FEATURE UNION

How to use the extracted global features and native features rationally to realize effective hippocampal interneuron characterization is considered as the next problem. If we just merge the two groups of features end-to-end by Array (vector) to generate a new feature vector, it will lack rational use of valuable identification data, feature union through effective optimization combination not only retain the participant multi-feature identification data, but also to some extent eliminate redundant data due to subjective factors, is of great significance to improve the characterization rate. In this paper, Euclidean Correlation Algorithm (ECA) is introduced to achieve feature union.

Euclidean Correlation Algorithm (ECA) process is as follows:

Step1, extract the global and native eigenvectors of the same Array based on the algorithms described in 2.1 and 2.2 to construct the transformed sample space A and B;

Step2, compute the overall covariance Array $S_{xx}$ and $S_{yy}$ of A and B segments, and mutual covariance Array $S_{xy}$;

Step3, according to Theorem 1 in the literature, compute the nonzero eigenvalues $\lambda_1^2 \geq \lambda_2^2 \text{L} \geq \lambda_r^2$ of $G1$ and $G2$ as well as the corresponding orthonormal eigenvectors $u_i, v_i (i = 1, 2, \text{L } r)$;

Step4, according to Theorem 1 to compute all the typical estimate vector $\alpha_i 与 \beta_i (i = 1, 2, \text{L } r)$, take the former $d$ pairs of the estimate vector to form the transposition Array $W_x$ and $W_y$

The above ECA estimate arrays are constructed with all the native feature subsets $\{F_i | i = 1, 2, \text{L}, 10\}$ and global features $F_A$ respectively,

Then 10 estimate Array pairs $[W_y^i, W_x^i](i=1,2,\text{L},L)$ are derived, and the estimate Array is used to respectively extract features from the corresponding feature subsets.

3.2 Classification Algorithm

For a certain sample to be charecterized, firstly, it is divided into $L$ pcs subsets according to the uniform rules, and then based on the construction method of the above sub-Arrays, $L$ pcs feature sub-vectors $(t^1, t^2, \text{L}, t^L)$ to be tested are derived, and the consistency between two feature sub-vectors $y^i$ and $t^i$ can be defined as:

$$consi(y^i, t^i) = -(y^i - t^i)W_y^i(W_y^i)^T(y^i - t^i) \tag{10}$$

First, compute the displacement between the *ith* sub-segments to be tested and the feature subvector used in the corresponding sub-Array, and then determine the categories of the sub-sample and a sub-sample most consistent with the sub-sample, call it as $C^i$. Finally, correct ultimate attribution of the sample to be tested by simple voting, and the probability that the sample to be tested is charecterized as the *kth* class can be defined as:

$$P_k = \frac{1}{L}\sum_{i=1}^{L} d^i \tag{11}$$

In the formula, when the first sub-mode of the sample to be tested is judged as the first *kth* class, the value of $d^i$ is 1, otherwise the value is 0. Finally, the characterization result of the sample under test is recorded as:

$$Iden = \arg\max(P_k) \tag{12}$$

# 4. EXPERIMENTAL RESULTS AND ALGORITHM

## 4.1 PRETREATMENT

In this paper, we test the hippocampal interneuron database, mainly including GMR, MIT, NCBR, MIT-CBCL, Fintech, KAIPO, UMASS, MITB, Grimace, UTMB, MUCT. According to the requirements of 2.1, Imprint size and data set were pre-processed, the main contents include:

For the GMR database, use the original 92 * 112 bmp imprint.

For the MIT database, use the original 100 * 100 bmp imprint.

For the NCBR database, as the 45th category is as same as the 51th category, the 46th category is as same as the 52th category, the 47th category is as same as the 53th category, the 48th category is as same as the 54th category, the 49th category is as same as the 55th category, the 50th category is as same as the 56th category, categories 51-56 are eliminated to prevent misclassification.

For the MIT-CBCL database, the original imprint is in the 115 * 115 pgm format. For easy calculation of TAT features, the size is normalized to 80 * 80 and converted to bmp format.

For the Fintech database, the hippocampal interneuron area is segmented using a given hippocampal interneuron area calibration point, then normalized to 80 * 100 size, and 19 categories are selected from which there are enough segments for characterization.

For the KAIPO database, use the 64 * 64 data provided by Deng Cai.

For UMASS database, the original database provides 92 * 112 hippocampal interneuron cut pgm format imprint.

For the MITB database, use the original Cropped MITB database. Since each category in the original database has a different number of segments, each category randomly selects 15 imprints for experiment, ensuring that each category has the same number of imprints.

For the grimace database, the original database 180 * 200 imprint contains too much background, cut the hippocampal interneuron area then normalize it to 80 * 80 size.

For the UTMB database, it is divided into 10 categories according to the nomenclature, and 20 segments randomly selected from each category are normalized to 80 * 80 for experiments.

For the MUCT database, there are 276 categories in the original database, 199 categories in the first (1-308), 15 segments in each, 77 categories (401-625) in the back, 10 segments in each, and the first 199 categories are used for the experiment: cut the hippocampal interneuron area according to provided landmark point, and normalize them to 80 * 80 for experiment.

| Name | Number of categories | Size | The sample number of each class | Train | Test |
|---|---|---|---|---|---|
| **GMR** | 40 | 92×112 | 10 | 7 | 3 |
| **MIT** | 15 | 100×100 | 11 | 8 | 3 |
| **NCBR** | 194 | 80×80 | 7 | 4 | 3 |
| **MIT-CBCL** | 10 | 80×80 | 200 | 10 | 190 |
| **Fintech** | 19 | 80×100 | 15 | 10 | 5 |
| KAIPO | 68 | 64×64 | 49 | 10 | 39 |
| UMASS | 20 | 92×112 | 7 | 4 | 3 |
| MITB | 38 | 84×96 | 59 | 10 | 49 |
| GRIMACE | 18 | 80×80 | 20 | 10 | 10 |
| UTMB | 10 | 80×80 | 20 | 10 | 10 |
| MUCT | 199 | 80×80 | 15 | 10 | 5 |

Table 1 Different Hippocampal interneuron Database Pretreatment Results

In selecting training / testing segments, it randomly extracts from each a specified number of segments from the sample banks for training / testing separately.

4.2 Overall performance test

This experiment compares the proposed method and the algorithm for dimensionality decline by using LDA on the different databases described in 4.1. Conduct TAT global feature mining according to 2.1, select No.5,6,7,8,10,11,13,14,15,16 sub-blocks as the native features for the corresponding treatment, and value mtree at 100, value mtry at different values respectively, do a number of experiments for the hippocampal interneuron database (change the value of the database and select different hippocampal interneuron banks). It mainly considers and discusses the required time for the identification and the correct rate of segments to test (percentage) and other indicators.

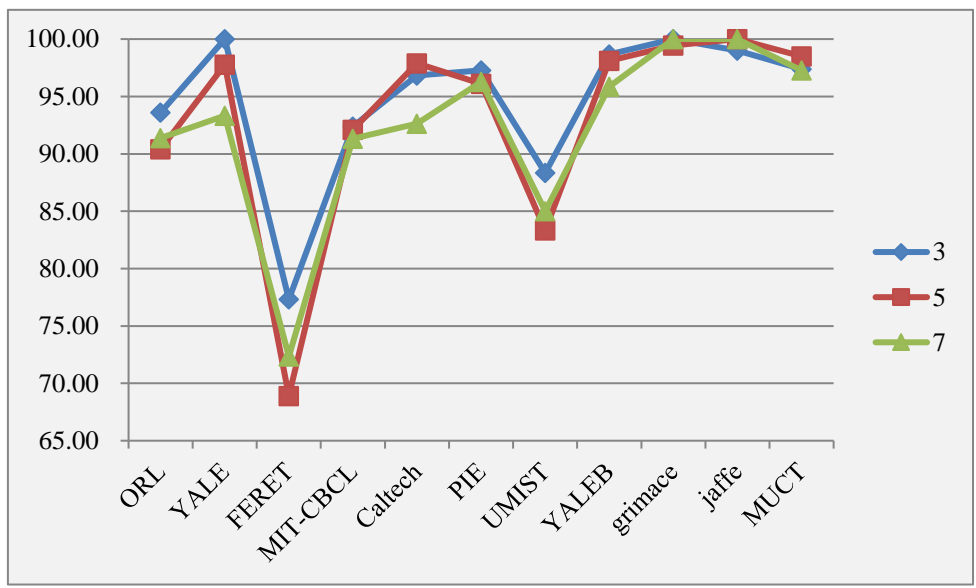

Figure 2. Effect of Different Test Segments on Characterization Accuracy

As can be seen from Figure2, reduce dimension by using the GPP algorithm, from the comprehensive comparison of different hippocampal interneuron database, when the number of segments at the value of 3, the test sample identification accuracy is higher overall, the hippocampal interneuron characterization accuracy in GMR, MIT, NCBR and other hippocampal interneuron banks is greater than that when the sample value of 5 or 7.

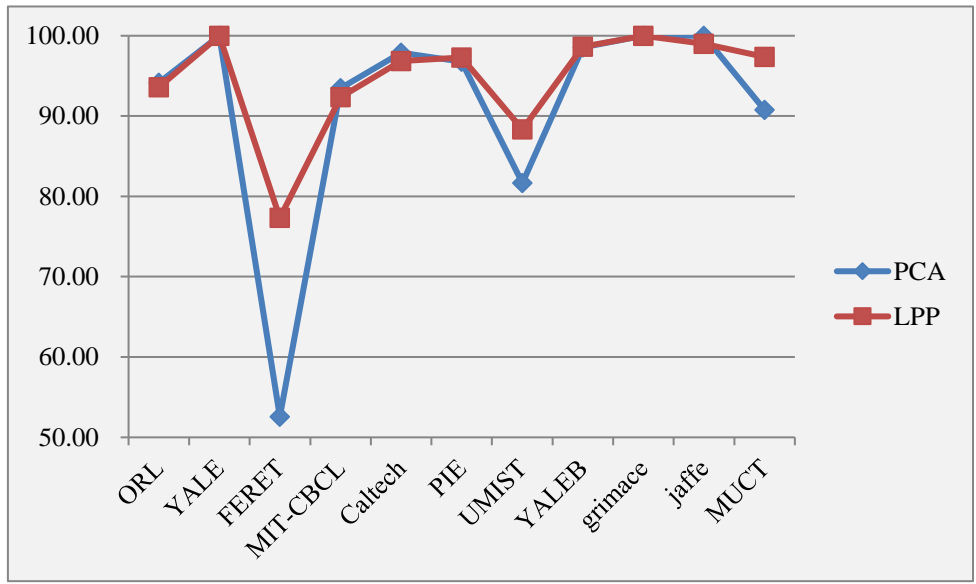

Figure 3. The Impact of Different Algorithms on Characterization Accuracy

In Figure 3, we fixed the test sample value to 3, and compared the accuracy of identifying the global features and native features by using PDA algorithm and GPP algorithm. It can be seen that after using the GPP algorithm to reduce the dimension, the correct rate of overall characterization is higher, especially in the NCBR, UMASS, MUCT database, the performance is obviously better than the use of FA dimensionality decline, while in the other database test results, the test accuracy is similar or close.

Table 2. Characterization Time Comparison

|  | GMR | MIT | NCBR | MIT-CBCL | Fintech | KAIPO | UMASS | MITB | grimace | UTMB | MUCT |
|---|---|---|---|---|---|---|---|---|---|---|---|
| FA | 1.04 | 0.14 | 9.25 | 0.05 | 0.11 | 2.45 | 0.15 | 0.55 | 0.10 | 0.04 | 32.78 |
| GPP | 1.09 | 0.07 | 2.72 | 0.04 | 0.12 | 3.52 | 0.07 | 0.52 | 0.11 | 0.04 | 6.86 |

It can be seen from Table 2 that based on the characterization time, GPP dimensionality decline is superior to FA algorithm in its entirety. This advantage is more obvious in NCBR and MUCT hippocampal interneuron database. In other hippocampal interneuron database, Characterization time is similar or close.

## 5. CONCLUSION

In this paper, the TAT feature descriptor is used to extract hippocampal interneuron features, and the global TAT feature and the native key part TAT feature are conducted union of feature layer to form the final classification feature, and the displacement similarity between the feature vectors is used for discrimination. Experiments conducted in multiple hippocampal interneuron databases show that the method using GPP dimensionality decline is better than using FA as a whole, which can meet the requirements of shortening the classification time and improving the characterization rate. In the future research, we also need to optimize the algorithm to further describe the degree of influence of different feature dimensions and key features on the characterization effect.